# Converting One-Way Streets to Two-Way Streets to Improve Transportation Network Efficiency and Reduce Vehicle Distance Traveled

Geoff Boeing[1] and William Riggs[2]
May 2022

**Abstract.** Planning scholars have identified economic, safety, and social benefits of converting one-way streets to two-way. Less is known about how conversions could impact vehicular distances traveled—of growing relevance in an era of fleet automation, electrification, and ride-hailing. We simulate such a conversion in San Francisco, California. We find that its current street network's average intra-city trip is about 1.7% longer than it would be with all two-way streets, corresponding to 27 million kilometers of annual surplus travel. As transportation technologies evolve, planners must consider different facets of network efficiency to align local policy and street design with sustainability and other societal goals.

## 1. Introduction

The past few years have witnessed much discussion of the benefits of converting one-way urban streets to two-way. Research has shown that two-way streets can improve network function (Gayah 2012; Gayah and Daganzo 2012) and be safer than one-way corridors (Riggs and Gilderbloom 2016). Urban designers have amplified this work to discuss the placemaking value of slower and safer streets that support multimodal travel. As a result, many cities are rethinking how urban streets should be designed and considering conversions of one-way corridors to promote safety and economic goals. This rethink involves renovating streets in urban neighborhoods that have suffered from years of neglect in favor of suburban development (Dover and Massengale 2013; Duany, Plater-Zyberk, and Speck 2001).

---

[1] University of Southern California
[2] University of San Francisco

The broad benefits of two-way conversions have gained traction in the planning literature, but narrower considerations of vehicular flow are more mixed. On one hand, one-way streets were originally designed to accelerate vehicles through automobile-oriented rights of way with simplified, efficient intersections. Yet there is more to efficiency than just this. One critical aspect is that, mathematically, the shortest path between two points in a fully bidirectional network will often be shorter than the shortest path between these same points in a network with some one-way streets—but the opposite is never true (Ortigosa, Gayah, and Menendez 2019). This is because one-way streets often force drivers to divert several blocks out of their way to overcome directionality restrictions to arrive at their destination.

Today, new mobility trends and technologies are changing travel patterns and behaviors in cities (Clewlow and Mishra 2017; Clewlow 2018; Riggs 2019; Millard-Ball 2019; Shaheen 2018). It is of critical importance to consider street network design in this context. Although vehicular electrification will ultimately mean decreased tailpipe emissions like carbon dioxide, nitrogen oxides, sulfur dioxide, and volatile organic compounds, it will not eliminate the substantial particulate matter emissions from rubber tire and brake dust (Thorpe and Harrison 2008; Harrison et al. 2012). Autonomous vehicles, mobility-as-a-service platforms, and shared fleets in tandem could also result in more continuous circulation of vehicles awaiting passengers. All told, distance-traveled measures of network efficiency become more important in certain emerging scenarios.

While planners have explored the safety and economic benefits of two-way street conversions, and engineers have explored their flow-efficiency impacts, an important research gap currently exists: it is not well-understood how these conversions could impact vehicle kilometers traveled (VKT). Theoretically, distances traveled could decrease as two-way streets enable more direct automobile routing, but to what extent is this the case empirically, given real-world network geometry and topology?

This article builds on recent theoretical work to investigate one-way to two-way street conversions' impacts on minimizing vehicular travel distance to better understand this aspect of transportation network efficiency. We construct two network models to simulate the efficiency of the drivable street network in the city of San Francisco. The first ("as-is") model preserves real-world one-way driving constraints, while the second ("as-proposed") model treats every street as bidirectional. Using California Household Travel Survey (CHTS) home and workplace coordinates to generate commute origins and destinations, we simulate distance-minimizing trips on each of the two networks. Two-way streets allow for significantly shorter average travel distances: the as-is network's average intra-city trip is about 1.7% longer than if it were all two-way streets, corresponding to 27 million kilometers of annual surplus travel just for intra-city trips.

As planners re-value and re-consider alternative uses for streets (Millard-Ball 2021), this suggests a new argument for street conversions beyond recent public health and economic development claims.

This article is organized as follows. The next section surveys the recent research literature around one-way/two-way street conversions. Then we explain the details of our study methods and present the empirical results. Finally, we discuss these findings and potential implications for transportation planning.

## 2. Background

### 2.1. The Logic of One-Way Streets

Nearly all city streets historically allowed bidirectional traffic. However, between the 1950s and 1980s, many cities converted two-way streets to one-way to promote traffic flow to burgeoning post-war suburbs for an increasingly automobile-centric, mid-century population (Appleyard 1980; Hall 1996; Handy, Paterson, and Butler 2003; Jackson 1987). The logic of these streets was centered around having centralized employment destinations with an inflow and outflow of commuters. In this vision of the city, one-way streets optimized high-capacity vehicular flow from one point to another, facilitating higher traffic volumes and centrifugal urbanization (Dover and Massengale 2013). City engineers designed roadways to maximize inflow and outflow, irrespective of distance, with hierarchies built for the emerging settlement patterns of the time—automobile-dependent suburban neighborhoods (Buchanan 1963; Bavarez and Newell 1967). This reengineering of streets enabled higher vehicular travel speeds over longer distances with less frequent stops (Hebbert 2005).

Heavily engineered streets optimized for vehicular traffic remain pervasive today. Yet they increasingly come into conflict with the other evolving goals of urban planning and design. A tension arises between humans moving quickly through cities and humans accessing and spending time in places (Hebbert 2005). City planners today increasingly emphasize compact development, density, environmental sustainability, and land use diversity (Stevens 2017; Salon et al. 2012). One-way—and reciprocally, two-way—streets play an important role in these debates.

### 2.2. The Argument for Two-Way Streets

The impacts of 20$^{th}$ century shifts in street design to promote automobile-oriented suburban expansion—and efforts to rethink them—have been the subject of substantial discussion in the recent planning literature. Researchers have found that these two-way to one-way conversions (and subsequent changes in adjacent land use) resulted in disinvestment and a degraded street-level human experience (Cervero and Kockelman 1997; Leinberger and Alfonzo 2012). A large body of work provides evidence that dense urban networks with a diversity of land uses promote more sustainable and diverse commute modes (Ewing and Cervero 2010; Frank et al. 2006; Frederick, Riggs, and Gilderbloom 2018). From the practitioner's perspective, Speck (2018) identified 78 cities that successfully restored one-way streets back to two-way, and at lower costs than anticipated. The financial feasibility of these projects is supported by a large body of work identifying the life-cycle cost effectiveness of roadway and traffic-calming changes (Noland et al. 2015).

The literature also identifies the potential of street conversions to support safety, equity, and economic goals (Riggs and Gilderbloom 2016, 2017; Riggs and Appleyard 2018). Instead of reinforcing automobile level-of-service (LOS) and vehicular flow as the ultimate aims of transportation planning, this body of research argues for the importance of a more comprehensive approach to urban quality of life. Changes to the right-of-way can support urban regeneration and revitalization. For instance, Riggs and Gilderbloom (2016) find that two-way streets create higher levels of economic activity compared to one-way streets because they expose businesses to individuals traveling in two directions versus those traveling in only one. Converting these one-way streets also creates an opportunity to repurpose lanes for other facilities (i.e., bikeways, wider sidewalks, parking, etc.) or dedicate some portion to micromobility and transit.

A parallel research topic has been network efficiency and safety. While the engineering literature tends to focus on first-order effects (such as LOS), equal attention should be paid to second-order effects. One-way streets may lead to higher speeds (Schneider, Grembek, and Braughton 2013), which may be desirable from an LOS or vehicular-throughput perspective, but other scholars have found that this is accompanied by an increased risk to pedestrian safety (Swift, Painter, and Goldstein 1998) and a false sense of security for drivers (Holahan 2013). Several studies have found that one-way streets are less safe than their two-way counterparts (Ewing, Schieber, and Zegeer 2003; Riggs and Gilderbloom 2016; Ewing and Dumbaugh 2009). Although some research suggests that older adults may struggle crossing two-way streets (Dommes, Cavallo, and Oxley 2013), broader longitudinal work suggests that one-way streets are less safe for children (Wazana et al. 2000): children are 2.5 times more likely to experience a pedestrian injury on a one-way versus two-way street. Gayah (2012) and Gayah and Daganzo (2012) argue that two-way streets make for more efficient networks and are safer for pedestrians because their complexity requires motorists to drive more attentively. They show that two-way streets are more efficient for traffic management, easier to navigate, and less confusing than one-way streets, especially for visitors to a new area. They also argue that vehicular-pedestrian conflict points are more easily anticipated on two-way streets.

**2.3. Re-Thinking Conversions' Effects**

Today, the research literature advocating for conversions back to two-way streets tends to focus on social and safety benefits. Meanwhile, the engineering literature demonstrating the superiority of one-way streets tends to focus on efficiency benefits, emphasizing signal timing and intersection flow. This is only part of the picture. Two-way conversions may offer a different set of network efficiency gains by reducing shortest-path distances along the network, which, in conjunction with signal timing and intersection efficiency, could influence total travel time, fuel consumption, and emissions.

Figure 1 presents a theoretical path model of how two-way street conversions impact vehicle travel, fuel consumption, and emissions. The two primary network policy and design inefficiencies are highlighted in yellow. Past research has focused on the right-side pathway, emphasizing one-way streets' positive benefits on signal timing and vehicular throughput. Less is known about the left-side pathway (delimited by the dashed gray bounding box) and its impact on VKT. Yet scholars argue that reducing VKT is critical to achieving environmental sustainability and climate goals (Barrington-Leigh and Millard-Ball 2017), particularly as we confront new mobility trends and technologies. Ortigosa, Gayah, and Menendez (2019) develop a theoretical model to argue that two-way networks could reduce VKT because they can provide drivers more direct routes from one location to another as one-way street networks increase the average driving distance. Although they can exhibit fewer stop-and-go effects at intersections due to signal timing, one-way street networks are less efficient from a distance-traveled standpoint. However, this argument makes a theoretical claim about idealized networks.

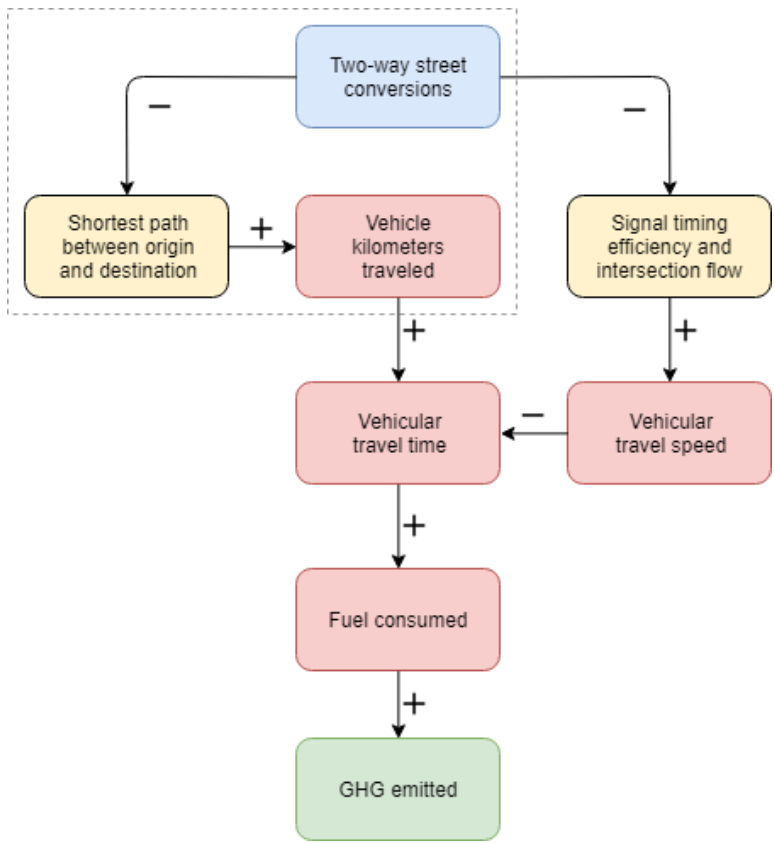

**Figure 1**. Theoretical model of one-way to two-way street conversions' effects on vehicle travel, fuel consumption, and emissions.

This article takes up this research gap and asks: to what extent do one-way restrictions increase shortest-path distances in a real-world city street network? Building our knowledge of Figure 1's left-side pathway, it advances the body of research on the efficiency of two-way networks by making two primary contributions. First, it builds on the theoretical model of Ortigosa, Gayah, and Menendez (2019) by conducting an empirical case study of San Francisco's street network. Second, it explores distance traveled through both a randomized set of origins/destinations (OD) and a survey-derived set to investigate network performance from different perspectives.

## 3. Methods

We construct two models of the city of San Francisco's drivable street network using data from OpenStreetMap. We choose San Francisco as a case study because it is reasonably large and reasonably representative of many US urban circulation systems: it contains a variety of neighborhoods with different network patterns and includes both one-way and two-way streets, like most large US cities. OpenStreetMap is a worldwide collaborative mapping project and platform (Boeing 2017). Its data are high-quality and detailed, providing a good source for modeling and analyzing transportation networks (Barron, Neis, and Zipf 2014; Zielstra, Hochmair, and Neis 2013; Christopher Barrington-Leigh and Millard-Ball 2017; Zhao et al. 2019). For each model, we download the drivable street network in the city of San Francisco using OSMnx, a Python-based toolkit for automatically downloading, modeling, and analyzing city street networks using OpenStreetMap data (Boeing 2018). We then filter out freeways to construct two models of the surface street network in the city.

The first model, $G_1$, preserves real-world one-way directionality constraints by modeling the network as a directed graph. The second model, $G_2$, treats every street as bidirectional using an undirected graph. Both graphs are topologically defined such that vertices represent intersections and dead-ends, and edges represent the street segments that link them.

Next, we construct two OD matrices to represent trip endpoints. The first matrix, $M_s$, uses the CHTS to generate an $n \times 2$ survey-derived OD matrix where $n$=1,133,333 daily vehicle trips that begin and end in San Francisco (San Francisco County Transportation Authority 2017). This is spatially representative of the actual origins (homes) and destinations (workplaces) for each surveyed household with both home and work within the city limits. For the second matrix, $M_r$, we randomly select $n$ origin and destination nodes from the graph to generate a randomized $n \times 2$ OD matrix. This second matrix provides a robustness test for the first to 1) corroborate that the simulated estimates are in the same range across both OD matrices, and 2) to more evenly cover different sections of the city than home-work commutes alone can.

Next, we simulate all the trips in each of the OD matrices on each of the two graphs, using two implementations of Dijkstra's shortest-path algorithm. The first algorithmic implementation, $A_1$, uses no weights and counts the topological graph distance (i.e., number of edges, which corresponds to the number of linear blocks) traversed on the shortest path between the origin and destination. The second implementation, $A_2$, weights edges by length, to minimize total distance traveled. Both assume free-flow travel without a queuing model. Finally, we use a dependent $t$-test for paired samples to analyze the differences in trips (for both OD matrices) between the two graphs to compare cumulative travel on the real-world network with one-way streets versus the alternative model with all bidirectional streets.

## 4. Results

First, we examine the number of blocks traversed given randomized origins and destinations. Using OD matrix $M_r$ and implementation $A_1$, we find that the typical (median) trip on $G_1$ must traverse 3 additional blocks (edges) compared to the typical trip on $G_2$ (51 edges versus 48 edges). We conduct a $t$-test and find the average (mean) trip traverses statistically-significantly more blocks on $G_1$ than on $G_2$: the difference in means is 2.75 blocks ($p$<0.0001).

Next, we examine the distance traveled given randomized origins and destinations. Using OD matrix $M_r$ and implementation $A_2$, the typical trip on $G_1$ travels a distance of 6,385 meters while the typical trip on $G_2$ travels 6,303 meters. Thus, the typical distance-minimizing trip on $G_1$ yields about 1.3% more VKT than that on $G_2$. We conduct a *t*-test and find the average trip is statistically-significantly longer on $G_1$ than on $G_2$: the difference in means is 80 meters ($p<0.0001$).

Next, we examine the number of blocks traversed given survey-derived origins and destinations. Using OD matrix $M_s$ and implementation $A_1$, we find that the typical trip on $G_1$ must traverse 2 additional blocks compared to the typical trip on $G_2$ (39 edges versus 37 edges). We conduct a *t*-test and find the average trip traverses statistically-significantly more blocks on $G_1$ than on $G_2$: the difference in means is 2.13 blocks ($p<0.0001$).

Finally, we examine the distance traveled given survey-derived origins and destinations. Using OD matrix $M_s$ and implementation $A_2$, the typical trip on $G_1$ travels a distance of 4,764 meters while the typical trip on $G_2$ travels 4,684 meters. Thus, the typical trip on $G_1$ yields about 1.7% more VKT than that on $G_2$. We conduct a *t*-test and find the average trip is statistically-significantly longer on $G_1$ than on $G_2$: the difference in means is 81 meters ($p<0.0001$).

The estimates produced by the survey-derived and randomized simulations demonstrate high consilience. Focusing on OD matrix $M_s$ (because it best reflects real-world parameterization), 86% of trips experience a decrease in distance traveled on the proposed alternative network with all bidirectional streets ($G_2$) relative to the as-is network with some one-way streets ($G_1$). For those trips that experience a decrease, the average is 95 meters. As noted earlier, the average trip on $G_1$ is about 1.7% longer than that of $G_2$. This difference is statistically significant but it is also practically significant. Using the San Francisco County Transportation Authority's figures for real-world daily VKT for trips that begin and end within the city (San Francisco County Transportation Authority 2017), we estimate that all else equal the presence of one-way streets increases daily VKT by more than 74,000 surplus kilometers and in turn increases annual VKT by over 27 million surplus kilometers.

**Table 1**. Results of the trip simulation.

|  | Survey-derived ODs ($M_s$) | | Randomized ODs ($M_r$) | |
| --- | --- | --- | --- | --- |
|  | $G_1$ | $G_2$ | $G_3$ | $G_4$ |
| Median trip length (m) | 4,764 | 4,684 | 6,385 | 6,303 |
| Mean trip length (m) | 5,012 | 4,930 | 6,518 | 6,439 |
| Median blocks traversed | 39 | 37 | 51 | 48 |
| Mean blocks traversed | 40.2 | 38.0 | 51.8 | 49.0 |
| Surplus annual VKT | 27,141,400 | – | 20,472,120 | – |

Note: the "surplus" row measures excess intra-city trip travel on the current network with one-way streets (G1) in relation to a fully-bidirectional alternative network (G2).

## 5. Discussion

Stevens (2017) argued that planners and engineers tend to overestimate the benefit of compact development, and that they "should probably assume for now that... [it] will have a small influence on driving... and municipal decision makers should not rely on compact development as their only strategy for reducing [VKT]." While this argument was critiqued by Handy (2017)—who argued that planning and engineering professionals cannot sufficiently reduce driving and address climate-related goals without compact development—our results suggest another arrow in planners' quiver to reduce VKT at the network scale.

Drivers may seek to minimize various factors when choosing a route, but the presence of one-way streets places a fundamental limit on the extent to which they have the ability to minimize VKT. While there are other factors like roadway types and traffic control that impact transportation network efficiency and travel psychology, the ability to optimize trips on a two-way versus one-way network can create network efficiencies and potential environmental benefits for society. Such conversions and signalization changes are usually less costly than expected—particularly when evaluated over the life-cycle of the street as an asset—and can bring ancillary economic development benefits (Speck 2018; Riggs and Gilderbloom 2016; Steuteville 2019; Noland et al. 2015).

If San Francisco converted its one-way streets to two-ways, the new network efficiencies could improve travel distance minimization by 1.7%. This allows us to sketch out a simple back-of-the-envelope *ceteris paribus* estimate of surplus fuel consumption and greenhouse gas emissions, assuming an average US fuel economy of 10.5 kilometers per liter and 2.4 kilograms of $CO_2$ released per liter of gasoline combusted. All else equal, a citywide two-way conversion policy in San Francisco could reduce annual VKT by 27 million kilometers, fuel consumption by 2.6 million liters, and carbon dioxide emissions by 6,200 metric tonnes (6.2 million kilograms)—just for intra-city trips that begin and end within San Francisco proper. The total impact of regional trips passing through the city would be larger still: CalTrans estimates that the total daily VKT of trips through the city is about 3.4 times greater than SFCTA's estimate of intra-city trips (CalTrans 2018). This is, however, a simple *ceteris paribus* analysis. In the real world, not all else is equal. A full accounting of conversions' impacts on fuel consumption and emissions would require simultaneous modeling of both the left and right pathways in Figure 1. This is beyond the goals of the present study, but the thought experiment opens the door for future research to incorporate integrated models of travel demand, route assignment, signalization, and congestion conditions to further explore the potential impacts of street network configuration on transportation systems.

This study has several practical outcomes for planning research and practice. These results provide a new and more nuanced understanding of the trade-offs that planners face in roadway design (i.e., prioritizing one-way vehicular efficiency at the cost of safety, distance traveled, fuel consumption, etc.). A predominant assumption over the past half-century has been that one-way streets offer more efficient vehicular travel. Our findings demonstrate that efficiency is in the eye of the beholder: one-way networks can offer signal timing and intersection efficiencies, but they produce other inefficiencies.

By forcing cars to traverse additional blocks, one-way streets entail an inherent, fundamental distance inefficiency. As urban transportation increasingly utilizes autonomous vehicles, electrification, and mobility as a service, these kinds of network efficiencies may become more important, particularly for autonomous vehicles that continuously circulate between passenger trips. Local travel can be more efficient if there is less circling distance and less routing around blocks unnecessarily to get to destinations. Even in a world where electrification eventually eliminates tailpipe emissions, longer distances traveled can still result in greater particulate matter pollution from rubber tire dust and brake dust.

In this study, we focused on the under-explored left-side pathway in Figure 1 to empirically estimate real-world impacts on VKT minimization. This preliminary analysis was conducted *ceteris paribus*—it did not allow for simultaneous changes in the right-side pathway. Limitations to extrapolation and generalizability include potential unknowns such an increase in idling and travel time based on intersection LOS as well as the slow but ongoing shift in fuel mix toward an electrified fleet. Traffic navigation apps also shift travel behavior in ways that are difficult to control for, and cities are now engaging in efforts to constrain use of networks by limiting the capacity of routing applications to use the entire street network. Moreover, little is known about the psychology of how or if network configurations impact travel decisions, or if on-the-whole they remain functions of trip purpose, convenience, and reliability which have all been dominant factors in the rise of the ridesharing economy and autonomous vehicles. Future research should conduct a detailed cost-benefit analysis, weighing conversion's benefits of safety, economic development, and reduced VKT and emissions against financial costs and vehicular LOS, which itself comes at a substantial social cost.

Future research should also consider the requirements of new mobility and model the various network distance and throughput effects through the multiple pathways of signal timing, intersection flow, and queuing that contribute to fuel consumption, emissions, and cost. Such an evaluation would be important not only in quantifying environmental trade-offs, but also potential productivity gains and time savings that ultimately could generate more economic value from street conversions. Finally, future research should expand these empirical findings by modeling and simulating two-way conversions' effects at multiple scales: neighborhood, municipality, and region. It should compare cities in different parts of the country to better understand the regional variation or importance of these policy and design changes.

## 6. Conclusion

Recent urban planning research has highlighted the social benefits of converting one-way streets to two-way streets. Yet much of today's road engineering research and practice instead emphasizes vehicular throughput via measures like LOS that preference one-way streets. This body of research has revealed much about signal timing and intersection flow, but little about distance-traveled measures of network efficiency that are important in certain emerging mobility scenarios.

This study addressed this gap by modeling San Francisco's street network and simulating trips along "as-is" versus fully-bidirectional model variants. We found that one-way streets fundamentally limit our ability to minimize VKT. If San Francisco's one-way streets were converted to two-way streets, intra-city trips could reduce distance traveled by nearly 1.7%, all else equal, corresponding to 27 million kilometers fewer miles traveled per year. With autonomous vehicles and continually circulating vehicles, the ability to realize more distance-efficient trips through network design may help lower economic and environmental costs.

Our findings provide planners and policymakers a new rationale for considering these street re-engineering projects. One-way to two-way conversions come with costs (financial, vehicular LOS, etc.), but they require a careful accounting of all social and environmental benefits. This study examined one under-explored aspect of this accounting. There may be additional political and engineering arguments for conversion projects beyond the recent public health and economic development arguments. While conversions have financial and political costs, they may help cities better prepare for sustainable future mobility by balancing various social benefits with network efficiency.

Swift, Peter, D. Painter, and M. Goldstein. 1998. "Residential Street Typology and Injury Accident Frequency." In *Presentation at the Congress for the New Urbanism VI, Denver CO*.

Thorpe, Alistair and Roy M. Harrison. 2008. "Sources and properties of non-exhaust particulate matter from road traffic: A review." *Science of The Total Environment* 400 (1–3): 270-282. https://doi.org/10.1016/j.scitotenv.2008.06.00

US Department of Energy. 2018. "How Can a Gallon of Gasoline Produce 20 Pounds of Carbon Dioxide?" *Fueleconomy.gov*. https://www.fueleconomy.gov/feg/contentIncludes/co2_inc.htm.

Wazana, Ashley, Vicki L. Rynard, Parminder Raina, Paul Krueger, and Larry W. Chambers. 2000. "Are Child Pedestrians at Increased Risk of Injury on One-Way Compared to Two-Way Streets?" *Canadian Journal of Public Health* 91 (3): 201–6. https://doi.org/10.1007/BF03404272.

Zhao, B, K Kumar, G Casey, and K Soga. 2019. "Agent-Based Model for City-Scale Traffic Simulation: A Case Study on San Francisco." In, 10. https://doi.org/https://doi.org/10.1680/icsic.64669.203.

Zielstra, Dennis, Hartwig H. Hochmair, and Pascal Neis. 2013. "Assessing the Effect of Data Imports on the Completeness of OpenStreetMap – A United States Case Study." *Transactions in GIS* 17 (3): 315–34. https://doi.org/10.1111/tgis.12037.

16